\documentclass[twocolumn,showpacs,amsmath,amstex,amssymb,mathfonts,prb]{revtex4}
\usepackage{graphicx,bm,units}

\begin{document}

\title{Robustness of the Spin-Chern number}

\author{Emil Prodan}
\address{Department of Physics, Yeshiva University, New York, NY 10016} 

\begin{abstract}
The Spin-Chern ($C_s$) was originally introduced on finite samples by imposing spin boundary conditions at the edges. This definition lead to confusing and contradictory statements. On one hand the original paper by Sheng and collaborators revealed robust properties of $C_s$ against disorder and certain deformations of the model and, on the other hand, several people pointed out that $C_s$ can change sign under special deformations that keep the bulk Hamiltonian gap open. Because of the later findings, the Spin-Chern number was dismissed as a true bulk topological invariant and now is viewed as something that describes the edge where the spin boundary conditions are imposed. In this paper, we define the Spin-Chern number directly in the thermodynamic limit, without using any boundary conditions. We demonstrate its quantization in the presence of strong disorder and we argue that $C_s$ is a true bulk topological invariant whose robustness against disorder and smooth deformations of the Hamiltonian have important physical consequences. The properties of the Spin-Chern number remain valid even when the time reversal invariance is broken.
\end{abstract}


\date{\today}

\maketitle

Quantum Spin-Hall insulators represent a new state of matter. They were predicted theoretically in Ref.~\onlinecite{Kane:2005ys} and soon after that they were found experimentally.\cite{B.A.-Bernevig:2006zr,Konig:2007uo} Samples made of such materials display dissipationless spin currents at the edges, that are robust against continuous deformations and disorder.\cite{Sheng:2005dz}  

The existence of the edge channels is due to the nontrivial topology of the bulk energy bands and two nontrivial topological invariants were proposed to describe this topology, virtually in the same time: the $Z_2$ invariant proposed by Kane and Mele \cite{Kane:2005vn} and the Spin-Chern number proposed by Sheng and collaborators \cite{Sheng:2006p324} (first mentioned in Ref.~\onlinecite{Sheng:2005dz}). In this paper we focus on the later invariant, which came under sustained scrutiny because it promised a finer classification of the Spin-Hall insulators. This was later argued not to be the case.\cite{Fu:2006p190,Fukui:2007sf}

The Spin-Chern number was computed by integrating the Berry curvature generated by imposing twisted boundary conditions on a finite sample. The numerical evidence given in Ref.~\onlinecite{Sheng:2006p324} implied that $C_s$ is a robust topological invariant. It was later observed, however, that one can continuously deform the model using spin rotations that keep the Hamiltonian's gap unchanged but switch the sign of the Spin-Chern number.\cite{Fu:2006p190,Fukui:2007sf} This argument shows that sometime the structure proposed in Ref.~\onlinecite{Sheng:2006p324} fails to be a smooth fiber bundle and that $C_s$ may not be well defined over the entire Spin-Hall zone of the phase diagram. The current understanding is that, whenever one crosses certain zones of the parameter space, $C_s$ jumps, but these jumps are always by an even number. Therefore, one can still use $C_s$ to formulate a $Z_2$ classification of the Spin-Hall insulators and to efficiently compute the $Z_2$ invariant. For this reason, the interest in the Spin-Chern number continues to be strong. For example, an efficient algorithm for numerical evaluations of $C_s$ was proposed by Fukui and Hatsugai,\cite{T:2007yg} and later the algorithm was used to map $C_s$ for aperiodic systems.\cite{Essin:wd} But other works totally dismiss the Spin-Chern number.\cite{Fu:2006p190,Fu:2007fu} For example Ref.~\onlinecite{Fu:2007fu} states that $C_s$ loses its meaning when the spin is not conserved.

In this paper we re-define the Spin-Chern number, this time directly for an infinite sample and without involving any boundary conditions. This shows, univocally, that $C_s$ is a topological invariant describing the bulk and not some edge as previous works implied. In our approach, the Spin-Chern number is defined as the half difference between the Chern numbers for the spin up and spin down sectors of the occupied space. We demonstrate that the spin up/down sectors can be rigorously defined even when $\hat{s}_z$ is not conserved. 

Furthermore, we state some very general results from the non-commuative theory of the Chern number\cite{Bellissard:1994rw} and then we show how to apply these results to demonstrate the quantization of the Spin-Chern number in the presence of strong disorder. Several physical consequences emerge from this analysis like the existence of critical energies below the Fermi level where the localization length diverges. Unlike the $Z_2$ invariant, the robust properties of the Spin-Chern number remain unchanged when the time reversal symmetry is broken. 

\section{Fiber bundles and their topological classification}

The classification of the topological insulators is intrinsically related to the classification of the fiber bundles. Intuitively, a fiber bundle is a collection of spaces, called fibers, that are indexed by a set of parameters that live on a manifold. The parameter space is called the base manifold and the reunion of all fibers is called the total space. An alternative view is to imagine the fibers as being attached to the points of the base manifold and then glued together by a prescribed topology. 

The fiber bundles encountered in condensed matter theory are generated by families of projectors $P_{\bm \theta}$, where ${\bm \theta}$'s live on a manifold ${\bm M}$ like the sphere, torus etc.. These projectors act on the same Hilbert space ${\cal H}$ and they generate the fibers ${\cal H}_{\bm \theta}$=$P_{\bm \theta}{\cal H}$. Since these fibers are embedded in the big Hilbert space ${\cal H}$, one can tell when two fibers  ${\cal H}_{\bm \theta}$ and ${\cal H}_{\bm \theta'}$ are close to each other and when they are not, therefore the bundle of fibers is already equipped with a topology.

Most of the fiber bundles studied in condensed matter are generated for finite samples by imposing parametrically dependent boundary conditions. To be more specific, consider a real space 2D lattice with $N$ quantum states per site and assume that a tight-binding Hamiltonian $H$ has been defined. The wave functions are $N$ dimensional complex vector fields defined over the finite lattice: ${\bm \Psi}(n_1,n_2)$, $n_1=1,\ldots, N_1$ and $n_2=1,\ldots,N_2$. Assume that we diagonalize $H$ by imposing boundary conditions like:
\begin{equation}\label{BoC}
\begin{array}{c}
{\bm \Psi}(N_1,n_2)=\hat{M}_1({\bm \theta}){\bm \Psi}(1,n_2) \medskip \\
{\bm \Psi}(n_1,N2)=\hat{M}_2({\bm \theta}){\bm \Psi}(n_1,1),
\end{array}
\end{equation}
where $\hat{M}_1({\bm \theta})$ and $\hat{M}_2({\bm \theta})$ are two $N$$\times$$N$ matrices depending parametrically on ${\bm \theta}$=$(\theta_1,\theta_2)$. Assume that, after the diagonalization, we find that the energy spectrum has a gap for all the allowed ${\bm \theta}$'s. In this case, we can construct a fiber bundle by simply considering the projectors $P_{\bm \theta}$ onto the states with energy below this gap. 

While this is a straightforward construction which pertains to easy numerical implementations, it has a few drawbacks which are very often overlooked. The possibilities of choosing the two matrices seem endless and same can be said about the parameter space. How can one then discern between a meaningful construction and the rest? The ultimate goal of these fiber bundle constructions is to characterize the bulk. Therefore, only a rigorous study of the thermodynamic limit can filter out the irrelevant matrices and the wrong parameter spaces. Most of the constructions will not survive in the thermodynamic limit either because the energy gap closes at some point or because the fibers seize to depend smoothly on the parameters. Unfortunately, studying the thermodynamic limit requires tedious numerical simulations, which most of the time can be carried only up to relatively small sample sizes. In fact, it is very rare when such studies are performed at all. We will discuss in the next section how to define fiber bundles directly in the thermodynamic limit.

We now turn to the problem of classifying the fiber bundles. If the dimension of the fibers ${\cal H}_{\bm \theta}$ is $N$, the fiber bundle generated by a family of projections is a U($N$) fiber bundle because two fibers ${\cal H}_{\bm \theta}$ and ${\cal H}_{\bm \theta'}$ can be linked by a unitary transformation $U_{\bm \theta \theta'}$. The question of how many topologically non-equivalent $U(N)$ fiber bundles can be build over a manifold ${\bm M}$ has been answered long time ago.\cite{Eilenberg:1953hq,Eilenberg:1954dw,Eilenberg:1954ay,Milnor:1956pt,Milnor:1956vf} It turns out that there is a universal manifold ${\bm X}$, called classifying space, such that all U($N$) fiber bundles over ${\bm M}$ can be generated from continuous functions $f:{\bm M}\rightarrow {\bm X}$, via the so called pull-back construction. Describing this pull-back construction is rather technical, but here is a very simple result: two fiber bundles over ${\bm M}$ generated by $f$ and $f'$ can be continuously deformed into each other if and only if the two functions $f$ and $f'$ can be continuously deformed into each other. Therefore, classifying the U($N$) fiber bundles over a manifold ${\bm M}$ is equivalent to classifying the continuous functions from ${\bm M}$ to ${\bm X}$. While a fundamental result in the  fiber bundle theory, this classification algorithm is somewhat formal because the space ${\bm X}$ is quite complicate [it is the infinite Grassmannian manifold $Gr_N({\bm C}^\infty)$].

A new development in the classification of low dimensional fiber bundles came from the work of Panati.\cite{Panati:2007sy} This work provides a computationally more manageable classifying approach and states, among other important things, that the fiber bundles over 2 dimensional manifolds without boundaries are completely classified by their Chern numbers. Note that the Chern number is just one invariant among many topologic invariants that can be associated with a fiber bundle. The Chern number can be explicitly computed by integrating the Berry curvature form:
\begin{equation}\label{BC}
dF = \mbox{Tr} \{P_{\bm \theta}[\partial_{\theta_1}P_{\bm \theta},\partial_{\theta_2}P_{\bm \theta}]\}d\theta_1 \wedge d \theta_2,
\end{equation}
 over the base manifold:
 \begin{equation}
C=\frac{1}{2\pi i}\int_{\bm M} dF.
\end{equation}
  The result of such integration is always an integer and this integer number is called the Chern number. 
  
Given Panati's result,\cite{Panati:2007sy} it is clear that in 2 dimensions there are no magic new topological numbers. Therefore, the Spin-Chern number must be related to the classical Chern number. As we shall see, this is indeed the case, but what is new is the structures for which the Chern number is computed.
 
\section{Working directly in the thermodynamic limit}  

There is one simple case in which one can work directly with infinite samples. This is the case of translational invariant Hamiltonians where one can use the Bloch fibration. The Bloch fibration is a unitary transformation from the Hilbert space of the infinite sample into a continuum direct sum of $N$ dimensional complex spaces:
\begin{equation}
\begin{array}{c}
U: {\cal H} \rightarrow \bigoplus_{{\bm k}\in {\cal T}} {\bm C}^N 
\medskip \\
(U{\bm \Psi})({\bm k}) = \frac{1}{2\pi} \sum_{\bm n} e^{-i {\bm n}{\bm k}} {\bm \Psi}({\bm n}),
\end{array}
\end{equation}
where ${\bm k}$ lives on the Brillouin torus ${\cal T}=[0,2\pi]$$\times$$[0,2\pi]$. $U$ transforms the original Hamiltonian in a direct sum of Bloch Hamiltonians:
\begin{equation}
UHU^{-1} = \bigoplus_{{\bm k}\in {\cal T}} H({\bm k}).
\end{equation}
Obtaining the explicit expressions of $H({\bm k})$ is, most of the time, a very simple exercise. Since $U$ is an isometry, if $H$ has an energy gap, then all $H({\bm k})$ have an energy gap.  We can then compute the projectors $P({\bm k})$ for the states below this gap and define the fiber bundle $\bigoplus_{{\bm k}\in {\cal T}} P({\bm k}) {\bm C}^N$,  with the Brillouin torus ${\cal T}$ as the base manifold. Consequently, we can compute a Chern number directly for the infinite sample. 

Consider now the case when the translational symmetry is lost, such as when disorder is present. For simplicity let us assume that the bulk gap remains open. If this is the case, we can define the total space simply as $P{\cal H}$, with $P$ being the projector onto the states below the gap, but how can we define a base manifold without the Brillouin torus? It turns out that the only way out of this dilemma is to define a non-commutative Brillouin torus, that is a torus whose points are operators instead of simple points on a manifold.

Describing this non-commutative torus is quite technical, but we can describe, in quite simple terms, how to define the Chern number. Let us show how this definition emerges from the translational invariant case. The Chern number computed as 
\begin{equation}
C=\frac{1}{2\pi i}\int_{\cal T} dF,
\end{equation}
 with the Berry curvature given in Eq.~\ref{BC} (where ${\bm k}$ takes the place of ${\bm \theta}$) has an expression in the real space:
\begin{equation}
C=2\pi i \ \mbox{tr}\{P\left [ [\hat{n}_1,P],[\hat{n}_2,P] \right ]\},
\end{equation}
where tr means trace over the quantum states at the origin, $[,]$ denotes the usual commutator and $\hat{{\bm n}}$ is the position operator: $(\hat{n}_{i}\Psi)(n_1,n_2)= n_{i}\Psi(n_1,n_2)$ ($i$=1,2). One can see that we don't really need the Brillouin torus to define the Chern number! Therefore, one could try to define the Chern number for the disordered case using this expression. The main question is, will this number be an integer? A fundamental result in non-commutative geometry says that this is indeed the case after averaging over the disorder.

We would like to state this result in more precise terms. For this we need to describe the disorder more precisely. To be specific, let us consider the addition of a random potential to $H$: 
\begin{equation}
V_\omega=\sum_{{\bm n},\alpha} \lambda_{\bm n}(\omega) |{\bf n},\alpha \rangle \langle {\bm n},\alpha|,
\end{equation}
where $\lambda_{\bf n}(\omega)$ is a random variable and $\alpha$ is the index of the quantum states ($\alpha$=$1,\ldots,N$). $\omega$ represents a particular disordered configuration and can be viewed as a point in the disorder configuration space $\Omega$, which must be equipped with a probability measure $d\mu(\omega)$ to be used for averaging over the disorder. We assume that the macroscopic state is translational invariant, in which case the lattice translations induce unitary transformations $u_{\bm n}$ on ${\cal H}$ and flows $t_{\bm n}$ on $\Omega$ such that $u_{\bm n}V_\omega u_{\bm n}^{-1}=V_{ t_{\bm n}\omega}$. The probability measure $d\mu(\omega)$ is assumed invariant and ergodic with respected to the flows $t_{\bm n}$. A concrete example will be the white noise, for which the disorder configuration space is the infinite cartesian product of $[-\frac{1}{2},\frac{1}{2}]$ intervals:
\begin{equation}
\Omega=\times _{{\bf n}\in {\bf Z}^2} \ [-\tfrac{1}{2},\tfrac{1}{2}].
\end{equation}
A point $\omega$ of this space is an infinite sequence:
\begin{equation}
\omega = (\ldots,\omega_{\bm n},\omega_{\bm n+1},\ldots)=\times_{{\bf n}\in {\bf Z}^2} \ \omega_{\bf n} 
\end{equation}
and $\lambda_{\bm n}(\omega)$ is taken as $\lambda_{\bm n}(\omega)= \omega_{\bm n}$. The probability measure $d \mu(\omega)$ is given by the infinite product of measures: $d\mu(\omega) = \times_{{\bf n}\in {\bf Z}^2} \ d\omega_{\bf n}$. The ergodic flow is simply $t_{\bf m} (\times_{{\bf n}\in {\bf Z}^2} \ \omega_{\bf n})= \times_{{\bf n}\in {\bf Z}^2} \ \omega_{{\bf n}-{\bf m}}$.

With these being said, consider a family of projectors $P_\omega$ acting on the whole Hilbert space ${\cal H}$.  We can now state the following rigorous result:\cite{Bellissard:1994rw} \medskip

\noindent {\bf Proposition 1.} If the matrix element $\langle {\bm n}|P_\omega|{\bm m}\rangle$ decays sufficiently fast (exponential decay is enough) with the separation $|{\bm n}-{\bm m}|$ and if $P_\omega$'s satisfy $u_{\bm n}P_\omega u_{\bm n}^{-1}=P_{ t_{\bm n}\omega}$, then:
\begin{equation}
C=2\pi i \int_\Omega d\mu(\omega)  \ \mbox{tr}\{P_\omega[[n_1,P_\omega],[n_2,P_\omega]]\}
\end{equation}
is an integer that is invariant to smooth deformations of $P_\omega$'s as long as they remain localized.\medskip

\noindent This is a very general results and it applies to any family of projectors and any type of disorder as long as the conditions inside the Proposition's statement are satisfied. 

If the projectors $P_\omega$ are associated to the spectral projectors onto the occupied states of a Hamiltonian, the Chern number defined above gives the disorder averaged Hall conductance. Proposition 1 tells us that this average remains quantized even in the presence of strong disorder. Since the Hall plateaus cannot be explained without the strong disorder, it was the above statement that completed our understanding of IQHE. 

\section{Defining the Spin-Chern number}

The Spin-Chern number was originally defined for a finite sample, by choosing $M_1({\bm \theta})$=$e^{i \theta_1}$ and $M_2({\bm \theta})$=$e^{2 i \theta_2 \hat{s}_z}$ in the boundary conditions Eq.~\ref{BoC}. Here we define the Spin-Chern number directly in the thermodynamic  limit.

Let us start our discussion from the concrete model of electrons in graphene:\cite{Kane:2005ys}
\begin{equation}\label{model}
\begin{array}{c}
H_0=-t\sum\limits_{\langle {\bm m \bm n} \rangle,\sigma} |{\bm m},\sigma\rangle \langle {\bm n},\sigma| \medskip \\
+i\lambda_{SO}\sum\limits_{\langle \langle {\bm m \bm n} \rangle \rangle,\sigma\sigma'}  [ {\bf s} \cdot ({\bf d}_{\bm k \bm m} \times {\bf d}_{\bm n \bm k} )]_{\sigma,\sigma'} |{\bm  m},\sigma\rangle \langle {\bm  n},\sigma'| \medskip \\
+i\lambda_R\sum\limits_{\langle {\bm m \bm n} \rangle,\sigma \sigma'}  [ \hat{{\bf z}}\cdot ({\bf s}\times {\bf d}_{\bm m \bm n})]_{\sigma,\sigma'} |{\bm n},\sigma\rangle \langle {\bm n},\sigma'|.
\end{array}
\end{equation}
Here, ${\bm m}$ and ${\bm n}$ denote the sites of the honeycomb lattice and $\sigma$ and $\sigma'$ the electron spin degrees of freedom, taking the values $\pm 1$. The Hamiltonian acts on the Hilbert space ${\cal H}$ spanned by the orthonormal basis $|{\bm n},\sigma\rangle$. The simple angular brackets denote the nearest neighbors and double angular brackets denote the second nearest neighbours. In the second sum, ${\bm k}$ represents the unique common nearest-neighbor of ${\bm m}$ and ${\bm n}$. The electrons are considered non-interacting. The three terms in Eq.~\ref{model} are the usual nearest neighbor hopping term, the intrinsic spin-orbit coupling preserving the lattice symmetries and the Rashba potential induced by the substrate supporting the graphene sheet.\cite{Kane:2005ys} We assume that the parameters in the Hamiltonian are chosen so that we are in the Spin-Hall zone of the phase diagram.\cite{Kane:2005vn} 

There are two sites per unit cell and two spin states per site, therefore the model has 4 states per unit cell. Let us neglect the disorder for a moment and let $P$ denote the spectral projector onto the states below the insulating gap of $H_0$. Under the Bloch fibration $U:{\cal H} \rightarrow \bigoplus_{{\bm k}\in {\cal T}} {\bm C}^4$, this projector becomes: $UPU^{-1}$=$\oplus_{{\bm k}\in {\cal T}}P({\bm k})$. The Bloch Hamiltonians $H({\bm k})$ display two  upper and two lower bands separated by an insulating gap. The Chern number of the projectors $P({\bm k})$ onto the lower (and for that matter also of the upper) bands is zero, as it will generically be for any time reversal invariant band model. According to Panati's result,\cite{Panati:2007sy} the fiber bundle of the occupied states is trivial, i.e. it has the simplest topological structure possible, that of ${\cal T}\times {\bm C}^2$. How can there be non-trivial topological structures? The answer is because in the Spin-Hall effect the non-trivial topological structures appear only in the spin sectors, when taken separately.  

We now show how one can define spin up and spin down sectors even for the case when $\hat{s}_z$ does not commute with the Hamiltonian. The spin operator is defined as 
\begin{equation} 
s_z |{\bm n},\sigma\rangle = \tfrac{1}{2}\sigma |{\bm n}, \sigma \rangle.
\end{equation}
 The goal is to slash the fiber bundle $\bigoplus_{{\bm k}\in {\cal T}}P({\bm k}){\bm C}^4$ into two non-trvial fiber bundles, that is to make the smooth decomposition $P({\bm k})$=$P_-({\bm k})$$\oplus$$P_+({\bm k})$ for all the ${\bm k}$'s of the Brillouin torus. At the first sight, this seems a very complicated job with no clear chances of success. The key idea is to use the operator $P\hat{s}_z P$ to do just that. Indeed, after the Bloch fibration, $UP\hat{s}_z PU^{-1}$ becomes $\oplus_{{\bm k}\in {\cal T}}P({\bm k})\hat{s}_z P({\bm k})$ and we can diagonalize each of the operators  $P({\bm k})\hat{s}_z P({\bm k})$. As we shall see, if the Rashba term doesn't exceed a threshold value, the spectrum of $P({\bm k})\hat{s}_z P({\bm k})$ consists of two isolated eigenvalues, positioned symmetrically and away from the origin, for all the ${\bm k}$'s of the Brillouin torus. The spectral projectors $P_\pm({\bm k})$ onto the negative/positive eigenvalues are smooth of ${\bm k}$ and can be use to achieve the decomposition $P({\bm k})=P_-({\bm k})\oplus P_+({\bm k})$. At this point we can can define the Chern numbers $C_\pm$ for the fiber bundles $\bigoplus_{{\bm k}\in {\cal T}}P_\pm({\bm k}){\bm C}^4$ via Eq.~\ref{BC} and define the Spin-Chern number as $C_s$=$\tfrac{1}{2}(C_+$$-$$C_-)$. This is our construction in a nutshell. 

We now step back and give a more general construction, without involving the Brillouin torus. Assume that the disorder is turned on and that we work now with the Hamiltonian $H(\omega)$=$H_0$+$\lambda V_\omega$. For simplicity, we assume that the amplitude of the disorder is not very large so that $H(\omega)$ still has an energy gap. Let $P(\omega)$ be the projector onto the states with energy below this gap and consider the operator $P(\omega)\hat{s}_z P(\omega)$. When the Rashba term is zero, $\hat{s}_z$ commutes with $H(\omega)$ and for that reason it also commutes with $P(\omega)$. Consequently, its spectrum consists of just two points, $\pm \tfrac{1}{2}$. These eigenvalues are highly degenerate. When the Rashba interaction is turned on, $\hat{s}_z$ and $P(\omega)$ no longer commute, therefore the degeneracy is lifted and the eigenvalues of $P(\omega)\hat{s}_z P(\omega)$ spread towards the origin. As we shall see, if the amplitude of the Rashba term does not exceed a threshold value, the spectrum of  $P(\omega)\hat{s}_z P(\omega)$ remain confined into two isolated islands, with the origin separating them. Therefore, we can define the spectral projectors onto the states with eigenvalues below the origin: $P_-(\omega)$, and above the origin: $P_+(\omega)$. To be able to define the non-commutative Chern numbers, we must demonstrate that the matrix elements $\langle {\bm m}|P_\pm(\omega)|{\bm n}\rangle$ decay exponentially fast with the separation $|{\bm m}-{\bm n}|$.

The exponential decay is required by the conditions stated in the Proposition 1. However, we would like to pause here and discuss what is the implication of this exponential decay for the translational invariant case. In this case, we have $P_\pm$=$\int  P_\pm({\bm k})d{\bm k}$ and it is known since the work of Kohn\cite{Kohn:1959p1023} that only smooth integrals of ${\bm k}$ lead to exponentially localized functions in real space. The reciprocal statement is also known to be true since the work of des Cloizeaux,\cite{Cloizeaux:1964cr,Cloizeaux:1964px} namely if a total projector is exponentially localized then its Bloch decomposition is smooth with ${\bm k}$. The smoothness of $P_\pm({\bm k})$ on ${\bm k}$ is essential because the Berry curvature involves derivatives over ${\bm k}$. Therefore, the exponential localization of $\langle {\bm m}|P_\pm(\omega)|{\bm n}\rangle$ is essential for both translational invariant and disordered cases. The good news is that exponential localization is always simpler to demonstrate than the smoothness with ${\bm k}$.

Once we demonstrate the exponential localization of the kernels, we can define the Chern numbers for the spin up and spin down sectors via:
\begin{equation}
C_\pm=2\pi i \int_\Omega d\mu(\omega)  \ \mbox{tr}\{P_\pm(\omega)[[\hat{n}_1,P_\pm(\omega)],[\hat{n}_2,P_\pm(\omega)]]\}
\end{equation}
Since $\hat{s}_z$ commutes with the translations, we have 
\begin{equation}
u_{\bm n}P_\pm(\omega) u_{\bm n}^{-1}=P_\pm ( t_{\bm n}\omega)
\end{equation}
 therefore, according to Proposition 1, $C_\pm$ are integers that do not change under continuous deformations as long as $P_\pm(\omega)$ remain exponentially localized. As we shall see, the exponential localization is guaranteed by the existence of the energy gap for $H(\omega)$ and by the spectral gap of $P(\omega)\hat{s}_z P(\omega)$. To compute the Spin-Chern number explicitly, we can turn off the disorder and the Rashba interaction, in which case the model Eq.~\ref{model} reduces to two decoupled Haldane models,\cite{Haldane:1988dz} for which $C_\pm$=$\pm1$ if $\lambda_{SO}>0$ and $C_\pm$=$\mp 1$ if $\lambda_{SO}<0$.

We now start the proof of the exponential localization of the projectors $P_\pm(\omega)$. To ease the notation we drop $\omega$. First, let us show that the gap $\Delta_1$ of the operator $P \hat{s}_z P$, viewed as an operator on the space of occupied states ${\cal K}$=$P{\cal H}$, remains open and clean when the Rashba term is turned on, as opposed to immediately closing or filling with additional spectrum due to some instability. For this, we notice first that the projector $P$ in the occupied space is analytic in $\lambda_R$. This property is protected by the insulating gap $\Delta_2$ of the Hamiltonian $H(\omega)$. We then show that, at least for small $\lambda_R$, $P \hat{s}_z P-\zeta I_{\cal K}$ is invertible for $\zeta$ in the vicinity of 0 [here $I_{\cal K}$ is the identity operator for the space ${\cal K}$]. One could try to work with the expression 
\begin{equation}
(P s_z P-\zeta I_{\cal K})^{-1}=P(P s_z P-\zeta I)^{-1}
\end{equation}
and use the continuity of $P$, but this expression has a problem when $\zeta$=0 since $(P \hat{s}_z P-\zeta I)^{-1}$ diverges because we include the un-occupied states where $P \hat{s}_z P$ is zero.

Here is an alternative approach inspired from Ref.~\onlinecite{Nenciu:1998cr}. Let $R(\zeta)=P(\hat{s}_z-\zeta)^{-1}P$, with $\zeta$ in a neighborhood of zero. We have successively:
\begin{equation}
\begin{array}{c}
R(\zeta)(P\hat{s}_z P- \zeta I_{\cal K}) = \medskip \\
P(\hat{s}_z-\zeta)^{-1}\{(\hat{s}_z-\zeta)P+[P,s_z]\}P \medskip \\
=P+P(\hat{s}_z-\zeta)^{-1}[P,\hat{s}_z]P\equiv I_{\cal K}+Q.
\end{array}
\end{equation}
$Q$ is small, at least for small $\lambda_R$, since
\begin{equation}
Q=P(\hat{s}_z-\zeta)^{-1}[P-P_0,\hat{s}_z]P,
\end{equation}
where $P_0$ is the projector onto the occupied states for $\lambda_R=0$. The small factor comes from $P-P_0$, which is proportional to $\lambda_R$. Above, we used the fact that $\hat{s}_z$ commutes with $P_0$, even in the presence of disorder. In this case, the operator $I_{\cal K}+Q$ has an inverse given by the convergent series $\sum_{n=0}^\infty (-Q)^n$ and we find:
\begin{equation}
(I_{\cal K}+Q)^{-1}R(\zeta)(P \hat{s}_z P- \zeta I_{\cal K}) =  I_{\cal K}, \ \text{or}
\end{equation}
\begin{equation}\label{resolvent}
(P \hat{s}_z P- \zeta I_{\cal K})^{-1}=(I_{\cal K}+Q)^{-1}P(\hat{s}_z-\zeta)^{-1}P.
\end{equation}
Thus we showed that resolvent of $P \hat{s}_z P$ is finite for $\zeta$ near the origin, which excludes the existence of any eigenvalues in this region.  

We now show that, as long as the gaps $\Delta_1$ and $\Delta_2$ of $P \hat{s}_z P$ and $H(\omega)$ remain opened, the spectral projectors $P_\pm$ are exponentially localized. By exponential localization of an operator $T$ we mean the existence of a strictly positive $\alpha$ such that
\begin{equation}
|\langle {\bm m},\sigma|T|{\bm n},\sigma'\rangle | \leq \text{ct.} \ e^{-\alpha |{\bm m} - {\bm n}|}.
\end{equation}
We will use the following simple observation: if $U_{\bm q}$ denotes the non-unitary transformation 
\begin{equation}
U_{\bm q}|{\bm n},\sigma \rangle = e^{{\bm q}\cdot {\bm n}} |{\bm n},\sigma\rangle,
\end{equation}
then the following is true.\cite{Prodan:2006cr} \medskip

\noindent {\bf Proposition 2.} If $T$ is exponentially localized, then $T_{\bm q}\equiv U_{\bm q}TU_{-\bf q}$ is a bounded operator for all orientations of ${\bm q}$, provided $|{\bm q}|$ is smaller than $\alpha$, and conversely: if $T_{\bm q}$ is bounded for any orientation of ${\bm q}$ and $|{\bm q}|<\alpha$, then $T$ is exponentially localized with a localization exponent equal or larger than $\alpha$. \medskip

\noindent Also, it is a fact that if $T$ is exponentially localized, then [$\| \ \|$ denotes the operator norm]
\begin{equation}\label{qq}
\|T-T_{\bm q}\| \rightarrow 0 \ \text{as} \ q\rightarrow 0,
\end{equation}
i.e. the difference between $T$ and $T_{\bm q}$ is small for $q$ small. 

\begin{figure}
 \center
 \includegraphics[width=8cm]{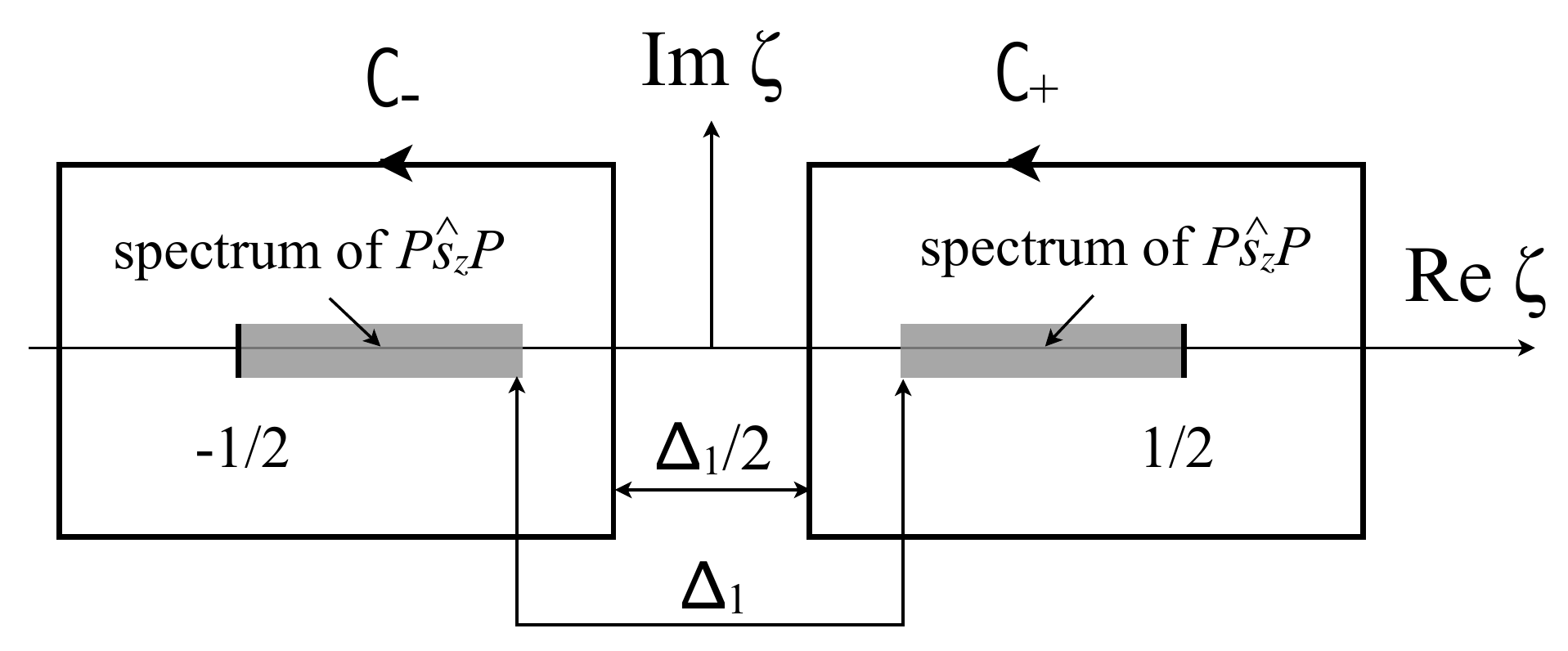}
 \caption{Illustration of the spectrum of $P\hat{s}_zP$ for small Rashba coupling, of the gap $\Delta_1$ and of the integration contours ${\cal C}_\pm$ used in the main text.}
 \label{fig}
\end{figure}

Since
\begin{equation}
P_\pm = \frac{i}{2\pi} \int_{{\cal C}_\pm} P(P \hat{s}_z P- \zeta I)^{-1} d\zeta,
\end{equation}
it is enough to show the exponential localization of $P(P \hat{s}_z P- \zeta I)^{-1}$ for all $\zeta$ on the contours ${\cal C}_\pm$ shown in Fig.~\ref{fig}. Now the projector $P$ itself is exponentially localized and a general proof of this statement can be found in Ref.~\onlinecite{Prodan:2008ai}. Then $P_{\bm q}$ is bounded and we have successively:
\begin{equation}\label{conv}
U_{\bm q} P(P \hat{s}_z P- \zeta I)^{-1}U_{\bm q}=P_{\bm q}(P_{\bm q} \hat{s}_z P_{\bm q}- \zeta I)^{-1},
\end{equation}
and
\begin{equation}
P_{\bm q} s_z P_{\bm q}- \zeta I=P s_z P- \zeta I+P_{\bm q} s_z P_{\bm q}-P s_z P
\end{equation}
thus
\begin{equation}
\|P_{\bm q} \hat{s}_z P_{\bm q}- \zeta I\|\geq \frac{\Delta}{2}-\|P_{\bm q} \hat{s}_z P_{\bm q}-P \hat{s}_z P\| 
\end{equation}
which is strictly positive for small $q$ due to Eq.~\ref{qq}. In other words, $P_{\bf q} s_z P_{\bf q}- \zeta I$ is invertible, at least for small $q$, which shows that all operators appearing in the right hand side of Eq.~\ref{conv} are bounded. Consequently, $P_\pm(\omega)$ are exponentially localized.

\section{Discussion}

Our results show that the Spin-Chern number is a topological invariant that is well defined in the thermodynamic limit. It is now clear that the robustness {\it is due to the existence of two spectral gaps}: the insulating gap of the Hamiltonian and the spectral gap of the operator $P\hat{s}_zP$. Contrary to many folkloristic believes, not all topological invariants are robust against disorder. In fact, to date, a rigorous proof for such statement exists only for the classic Chern number. The proof makes heavy use of the particular expression of the Chern number, thus it cannot be automatically extended to other topological invariants. Therefore, it is remarkable that we can add the Spin-Chern number to the list of topological invariants that are robust to disorder.

The robustness of the Spin-Chern number has physical consequences. We recall that, according to Ref.~\onlinecite{Abrahams:1979et} and \onlinecite{Anderson:1980bu}, generically for dimension smaller than two, all states are exponentially localized at any strength of disorder; for dimension greater than two, there exist extended states for low disorder and for dimension two all the states are localized except for states corresponding to isolated critical energies at which the localization length diverges. The Chern number in IQHE not only has a classifying role, but its quantization in the presence of disorder implies the existence of such critical energy regions below the Fermi level.\cite{Bellissard:1994rw} The quantization of the Spin-Chern number leads to the same conclusion, which explains the finite Spin-Hall conductance in the presence of strong disorder.\cite{Sheng:2005dz} This remarkable property is solely related to the spin, because the Spin-Chern number is robust against perturbations that violate the the time reversal symmetry. Indeed, none of our constructions or proofs use the inversion symmetry as an argument.

The operator $\hat{s}_z$ is not special. In fact, $C_s$ remain invariant to deformations of the Hamiltonian {\it and} of $P\hat{s}_zP$, as long as the spectral gaps of the two operators remain open. Closing any of the two gaps can result in jumps for the the Spin-Chern number. The jumps are always by an even number. The argument against the Spin-Chern number in Refs.~\onlinecite{Fu:2006p190} and \onlinecite{Fukui:2007sf} was that, using rotations in the spin sector, one can connect the Hamiltonians with $+\lambda_{SO}$ and with $-\lambda_{SO}$, without changing the insulating gap. But at the end of such rotation, $C_s$ changes sign and it was concluded that $C_s$ is not a well defined topological invariant. The change of sign was attributed to the closing of the gap of the Hamiltonian with the spin boundary conditions. We now can give an alternative explanation: during the rotation in the spin sector, the gap of $P \hat{s}_z P$ closes and then opens again. This problem can be easily fixed, the solution being to deform not only the Hamiltonian but also the operator $P \hat{s}_z P$. For the case of spin rotations, this can be easily accomplished by applying the spin rotations to $\hat{s}_z$ inside $P \hat{s}_z P$. If there is such easy fix, then the old question resurfaces: does the Spin-Chern number contain more information than the $Z_2$ invariant? The answer is no. After the continuous rotation mentioned above, $P \hat{s}_z P \rightarrow -P \hat{s}_z P$ and $P_\pm \rightarrow P_\mp$. Thus, there is no canonical way to chose the projectorss $P_\pm$ and, since this choice determines the sign of $C_s$, the sign contains no additional information.

The construction can be easily generalized. For more complex insulators, it is very probable that one can replace $\hat{s}_z$ by other non-trivial operators $\hat{w}$ by combining spin and/or point group symmetry operators. Using such operators one might discover nontrivial topological structures in seemingly trivial insulators. For example, it was recently pointed out \cite{Kuge:2008rq} that certain surface states in ordinary semiconductors can have topological origins. The non-zero Chern numbers for the spectral sectors of $P\hat{w}P$ will then lead to a $\hat{w}$-Hall effect. For other models, we can find operators $P\hat{w}P$ that have $n$ islands of isolated spectrum with exponentially localized projectors. If $\hat{w}$ commutes with the translations of the unit cell, we can define a Chern number for each spectral island of $P\hat{w}P$. A simple example of this type is the model of Eq.~\ref{model} with spin $\frac{3}{2}$ particles. For $\lambda_R$=0, $\hat{s}_z$ commutes with the Hamiltonian and $P{\cal H}$ splits into four sectors, corresponding to $s_z=-\frac{3}{2}$, $-\frac{1}{2}$, $\frac{1}{2}$ and $\frac{3}{2}$. Turning on the disorder will not affect these eigenvalues. For each sector we can define a Cern number, which take the values (assuming $\lambda_{SO}>0$): $C_{-\frac{3}{2}}=C_{-\frac{1}{2}}=-1$ and $C_{\frac{1}{2}}=C_{\frac{3}{2}}=1$. When the Rashba interaction is turned on, the spectrum of $P\hat{s}_zP$ spreads, but is still contained into four isolated spectral islands. Thus, we can still split $P{\cal H}$ into four sectors and our analysis shows that the Chern numbers for each sectors will be conserved as long as the the insulating gap remains open and the spectral islands of $P\hat{s}_zP$ remain isolated. The spin-Chern number can be defined in various ways, depending how we group the sectors. If we repeat the construction for spin $\frac{1}{2}$ and put the negative $s_z$ sectors into ${\cal K}_-$ and the positive $s_z$ sectors into ${\cal K}_+$, $C_s$ becomes 2. Other possible groupings gives $C_s=0$ and $C_s=-2$. 

The existence of the nontrivial Chern numbers for different sectors does not automatically imply the existence of chiral edge modes. Generically, only the insulators with odd $C_s$ display such edge modes.\cite{Fu:2006p190} The discussion given here hardly touches the problem of classification and of the edge modes. For these two problems, one has to explore how the occupied states relate (connect) to the un-occupied states. This has to be done on solvable models. It is at this point where our analysis becomes relevant because now we have a guiding principle which tells when the realistic models, which should include disorder and will generally not be solvable, can be deformed into smooth solvable models without changing the topological invariants. The robustness of the edge modes was also recently investigated in Ref.~\onlinecite{Prodan:2009ud}, which introduced a quantized edge index. Although this edge index was also constructed via a splitting, the connection between the edge index and $C_s$ is not clear to us at this moment. 

We want to make a final remark about the edge modes. The existence of nontrivial topological sectors can indicate reach edge and surface physics. For example, for the model of spin $\frac{3}{2}$ particles, when cutting an edge we can be sure that four [which become 2] edge bands shoot out of the bulk spectrum. The bands hybridize and return back into the same part of the bulk spectrum were they originated. Nevertheless, these bands lead to [gapped] edge states, which can still be useful for practical applications. In the presence of disorder, these edge bands will localize, but if the localization length is large, these edge states, for example, can efficiently trap light and thus be useful in photovoltaic devices.

\section{Conclusions}

We have demonstrated that the Spin-Chern number can be defined in the thermodynamic limit without the need of any boundary conditions, therefore showing that the Spin-Chern number describes the bulk and not the edges as it is currently believed. By making a connection with the non-commutative theory of Chern number, we were able to demonstrate the robustness of the Spin-Chern number against disorder and smooth deformations of the models. The robustness, which is solely related to the spin and has nothing to do with the time reversal symmetry, has physical consequences: it implies the existence of critical energies below the Fermi level where the localization length diverges. We have also theorized possible generalizations of our construction, which could aid the search and discovery of new topological phases.

\noindent {\bf Aknowledgement.} This work was supported by an award from the Research Corporation.

\end{document}